\documentclass{appolb}

\usepackage{graphicx}
\usepackage{amssymb}
\usepackage{hyperref}
\usepackage[utf8]{inputenc}

\begin{document}

\title{Fluctuations of flow harmonics in Pb+Pb collisions at $\sqrt{s_{NN}}=2.76$~TeV from the Glauber model%
\thanks{e-mails: Maciej.Rybczynski@ujk.edu.pl, Wojciech.Broniowski@ifj.edu.pl}}

\author{Maciej Rybczy\'nski$^{1}$ and Wojciech Broniowski$^{1,2}$ 
\address{$^{1}$Institute of Physics, Jan Kochanowski University, PL-25406~Kielce, Poland}
\address{$^{2}$The H. Niewodnicza\'nski Institute of Nuclear Physics, Polish Academy of Sciences, PL-31342 Krak\'ow, Poland}}

\maketitle

\begin{abstract}
In the framework of the Glauber model as implemented in {\tt GLISSANDO~2}, we study the fluctuations of flow harmonics in Pb+Pb collisions 
at the LHC energy of $\sqrt{s_{NN}}=2.76$~TeV. The model with wounded nucleons and the admixture of binary collisions leads to 
reasonable agreement for the ellipticity and triangularity fluctuations with the experimental data from the ATLAS, ALICE, and CMS collaborations, verifying the assumption that the initial eccentricity is 
approximately proportional 
to the harmonic flow of charged particles. While the agreement, in particular at the level of event-by-event distributions of eccentricities/flow coefficients 
in not perfect, it leads to a fair (at the level of a few percent for all centralities except the most peripheral collisions) description
of the scaled standard deviation and the $F$ measure which involves the four-particle cumulants. We also discuss the case of quadrangular flow.  
Computer scripts that generate our results from the {\tt GLISSANDO 2} simulations are provided.
\end{abstract}

\PACS{25.75.-q}
  
\section{Introduction}

Studies of correlations and fluctuations are in the core of the heavy-ion physics program, 
as they carry valuable information on the dynamics of the system in the early and intermediate stages of the collision. 
In particular, the azimuthal angle distributions of the produced  hadrons have been a subject of extensive 
experimental studies at RHIC (see, e.g.,~\cite{Sorensen:2006nw,Alver:2007zz,Alver:2007rm}
and more recently at the LHC~\cite{Aamodt:2010pa,Abelev:2012di,Aad:2013xma,Aad:2014vba,Chatrchyan:2013kba}.

In this paper we present predictions of the Glauber model of the initial stage of the heavy-ion reactions for fluctuations of the flow harmonics. 
While many such studies have been presented in the literature (see, e.g., the references in the recent review in Ref.~\cite{Luzum:2013yya}, 
or the latest works~\cite{Qiu:2011iv,Niemi:2012aj,Bzdak:2013rya,Renk:2014jja,Fu:2015wba,Bravina:2015sda}), recently some confusion has been raised by results published in 
Ref.~\cite{Aad:2013xma,Aad:2014vba} claiming, that the Glauber model fails badly for the fluctuations of flow (cf. Figs.~13-14 from Ref.~\cite{Aad:2014vba} and Fig.~18 from Ref.~\cite{Aad:2013xma}). 
In this paper we show, that this is not the case, and that the agreement with the experimental results of Refs.~\cite{Aamodt:2010pa,Aad:2014vba,Chatrchyan:2013kba} holds at the expected level for a wide range of centralities.  
While the agreement is not perfect for the event-by-event distributions of eccentricities/flow coefficients, showing differences in the tails of these distributions, 
the global measures, such as the event-by event standard deviation or the $F$ measures, are reproduced at a level of a few percent for all centralities except the most peripheral collisions.
So the basic outcome of our work is that the Glauber model works for the description of the fluctuations of ellipticity and triangularity.
It also works, for a somewhat lesser accuracy, for  the  quadrangular flow.

The consistency of the Glauber model with the data is important, as the approach is used as one of the baselines of the early-stage modeling in numerous analyses, 
also the experimental ones, where the connection of centrality to the number of participants is made with the help of Glauber simulations. 
We provide computer scripts that generate our results from the {\tt GLISSANDO~2}~\cite{Broniowski:2007nz,Rybczynski:2013yba} simulations if the reader wishes to effortlessly repeat or extend our results.

The formalism used in this paper is described in detail in Ref.~\cite{Broniowski:2007ft}. In particular, all details concerning the statistical methods and the popular variants of the Glauber models may be found there, so in this paper we limit the presentation to the minimum.

\section{Glauber model}

We use {\tt GLISSANDO~2}~\cite{Broniowski:2007nz,Rybczynski:2013yba} to analyze two variants of the Glauber model with Monte Carlo simulations:
\begin{enumerate}

\item The {\it mixed} model, amending wounded nucleons~\cite{Bialas:1976ed} with an admixture of binary collisions~\cite{Kharzeev:2000ph,SchaffnerBielich:2001qj,Back:2001xy,Back:2004dy} in the 
proportion $\alpha$. The successful fits to particle multiplicities (see Ref.~\cite{Back:2004dy}) give $\alpha = 0.145$ at $\sqrt{s_{NN}}=200$~GeV. 
For the LHC energy of $\sqrt{s_{NN}}=2.76$~TeV we take $\alpha=0.15$. 

\item Each source from the mixed model may deposit entropy with a certain distribution of strength. 
Therefore, we superpose the Gamma distribution  over the distribution of sources and label this model. 
The choice of this distribution follow from the fact that when folded with the Poisson distribution for the production of  the number of 
particles at freeze-out, it yields the popular negative binomial distribution. 

\end{enumerate}
The expression for the initial entropy distribution in the transverse plane is
\begin{eqnarray}
s({\bf x}_T) = {\rm const} \left ( \frac{1-\alpha}{2} \sum_{i=1}^{N_{\rm w}} w_i g_i({\bf x}_T) + \alpha  \sum_{j=1}^{N_{\rm bin}} w_j g_j({\bf x}_T) \right ), \label{eq:gl}
\end{eqnarray}
where 
\begin{eqnarray}
g_k({\bf x}_T) = \exp \left( \frac{({\bf x}_T-{\bf x}_{T,k})^2}{2\sigma^2} \right) 
\end{eqnarray} 
describes the smearing of the sources (wounded nucleons or binary collisions) located at ${\bf x}_{T,k})$. The smearing parameter 
is $\sigma=0.4$~fm~\cite{Bozek:2013uha}. The center of the binary-collision source is at the mean of the location of the centers of the colliding nucleons.

The choice of weights $w_k$ requires a careful discussion. In our approach~\cite{Olszewski:2013qwa} there are two sources of fluctuations: 
in the early stage, stemming from the statistical nature of the collision process, and in the final stage, from statistical hadronization. 
When we are interested in the initial shape, we should include only fluctuations generated at this stage. With no weight fluctuations here we 
simply set $w_i=1$, and with the Gamma fluctuations included, $w_i$ are generated randomly from the Gamma distribution~\cite{Broniowski:2007nz}.

When we are interested in the multiplicity fluctuations (as in Sec.~\ref{sec:mult}), we look at the system in the final phase, therefore we need to overlay 
the Poisson distribution from the statistical hadronization. Then $w_i$ is generated from the Poisson distribution in the model with no weight fluctuations in the Glauber phase, 
and from the negative binomial distribution in the model with Gamma fluctuations in that phase.

\section{Multiplicity fluctuations in p+Pb collisions \label{sec:mult}}

\begin{figure}[tb]
\centerline{%
\includegraphics[width=0.6\textwidth]{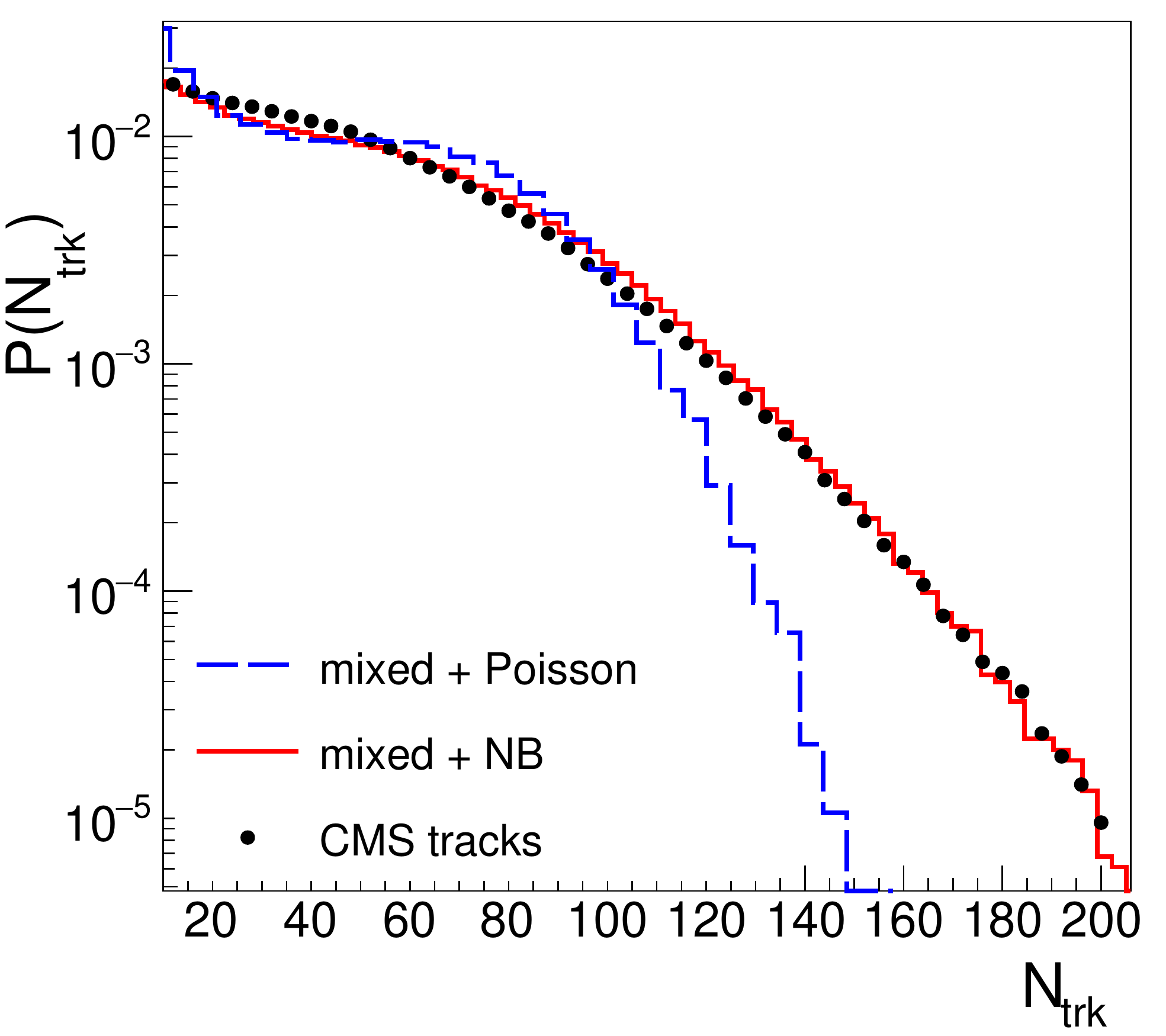}}
\caption{Multiplicity distribution in p+Pb collisions, where the CMS data~\cite{cmswiki}
are for charged tracks with $p_T > 0.4$~GeV and $|\eta| < 2.4$, and the lines denote {\tt GLISSANDO~2} results for 
the mixed+Poisson (dashed line) and mixed+NB (solid line) models. We note that the weight fluctuations in the Glauber phase (generated with the Gamma 
distribution) are essential for the agreement of the high-multiplicity tail.
\label{Fig:mult}}
\end{figure}

The need for the overlaid distribution comes from the physical fact that the individual collisions deposit fluctuating amount of energy in the transverse plane.  
Moreover, such fluctuations are necessary to reproduce the particle spectrum at very large multiplicities. One may use the  multiplicity distribution measured in the p+Pb collisions 
\cite{cmswiki} to adjust the parameters of the negative binomial distribution
\begin{equation}
N_{\lambda,\kappa}(n)=\frac{\Gamma(n+\kappa)\lambda^n\kappa^\kappa}{\Gamma(\kappa)n! (\lambda+\kappa)^{n+\kappa}}\ , \label{eq:NB}
\end{equation}
where the multiplicity $n~(=w_k)$ has the mean $\langle n \rangle = \lambda$ and variance $\sigma(n)^2 = \lambda(1+\lambda/\kappa)$. 
With $\lambda=9.3$ and $\sigma(n)=10.3$ we obtain the matching displayed in
Fig.~\ref{Fig:mult}, where we compare  the model result to the CMS data~\cite{cmswiki}. 
On the other hand, the model without the weight fluctuation in the early phase (labeled {\em mixed+Poisson}) in Fig.~\ref{Fig:mult} fails to reproduce the data.

That way we fix the parameters of the overlaid distribution. Correspondingly, in the analysis of the eccentricities in the next sections we use value of $\nu=0.9$ in the 
Gamma distribution
\begin{eqnarray}
g(w,\nu)=\frac{w^{\nu-1}\nu^\nu \exp(-\nu w)}{\Gamma(\nu)}, \;\; w \in 
[0,\infty). \label{gamma}
\end{eqnarray}
which yields $\langle w \rangle=1$ and $\sigma(w)=1/\sqrt{\nu}=1.054$. 

The realistic nucleon-nucleon inelastic collision profile for the LHC energies is taken from Ref.~\cite{Rybczynski:2013mla}. We use an excluded distance $d=0.9$~fm when 
generating the nucleon configurations in the nuclei. In the case of {\em mixed+$\Gamma$} variant we use the correlated configurations of nucleons in Pb nuclei provided 
by~\cite{Alvioli:2009ab}. The total inelastic nucleon-nucleon cross section is equal to $64$~mb for the investigated Pb+Pb collisions at $\sqrt{s_{NN}}=2.76$~TeV.

\section{Fluctuations of elliptic and triangular flow}

Due to collectivity of the fireball evolution, the azimuthal anisotropy of hadrons produced in the final state reflects  the initial spatial asymmetry of
the fireball in the transverse plane, which is due to geometry~\cite{Ollitrault:1992bk} and event-by-event 
fluctuations~\cite{Alver:2006wh,Voloshin:2006gz,Hama:2007dq,Alver:2010gr,Luzum:2011mm,Bhalerao:2014xra}. The observed particle distributions are characterized by the harmonic flow coefficients $v_{n}$, defined as the Fourier coefficients of the expansion
\begin{equation}
\frac{dN}{ d\phi} = \frac{N}{ 2 \pi}\left[ 1+2\sum_{n=2}^\infty v_n 
\cos\left[ n(\phi-\Psi_n)\right] \right].
\label{eq:vn}
\end{equation}
(in this paper we use the $v_n$ coefficients integrated over the transverse momentum for symmetric systems and at mid-rapidity).

Analogously, the eccentricity coefficients $\epsilon_n$ parametrize the shape of the initial fireball, and are defined in a given event as
\begin{equation}
\epsilon_{n} e^{i n \Phi_n}=\frac{\int dx_T s({\bf x}_T) \rho^n e^{i n \phi}}{\int dx_T s({\bf x}_T) \rho^n},
\end{equation}
where $\rho$ and $\phi$ are the polar coordinates corresponding to ${\bf x}_T$, and the source density $s({\bf x}_T)$ is given in Eq.~(\ref{eq:gl}).
The event-plane angles $\Psi_n$ and $\Phi_n$ are interesting in their own right~\cite{Bozek:2015bna, Bozek:2010vz}, but are not important for the analysis shown in this paper.

It has been argued (see, e.g., Ref.~\cite{Niemi:2012aj,Bzdak:2013rya,Fu:2015wba,Bozek:2014cva}) that to a good accuracy one has the proportionality 
(the ``shape-flow'' transmutation)
\begin{eqnarray}
v_n=\kappa_n \epsilon_n, \;\;n=2,3, 
\label{eq:linear}
\end{eqnarray}
where the constants $\kappa_n$ depend on features of the colliding system (centrality selection, mass numbers, collision energy) and the properties of the dynamics (viscosity of quark-gluon plasma, initial time of collective evolution, freeze-out conditions). 
Yet, when the above-mentioned conditions are fixed, Eq.~(\ref{eq:linear}) holds to sufficient accuracy that allows for model-independent predictions. Technically, Eq.~(\ref{eq:linear}) means that the response of the system to small shape deformation, for a given class on initial conditions such as centrality selection, is linear. The feature is limited to $n=2, 3$, as for higher harmonics nonlinear effects may be substantial~\cite{Teaney:2010vd}. 

\begin{figure}[tb]
\begin{center}
\includegraphics[width=0.51\textwidth]{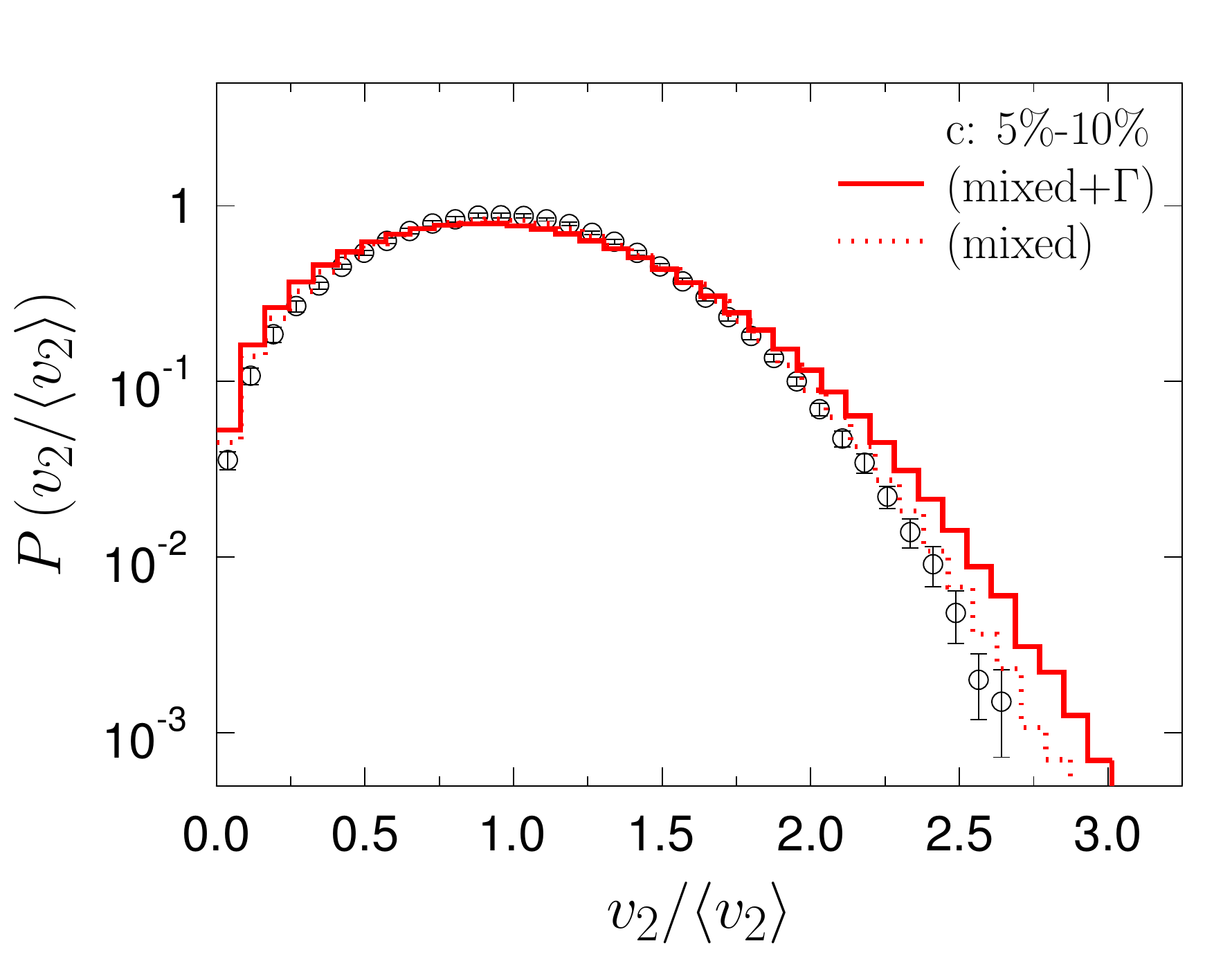}\includegraphics[width=0.51\textwidth]{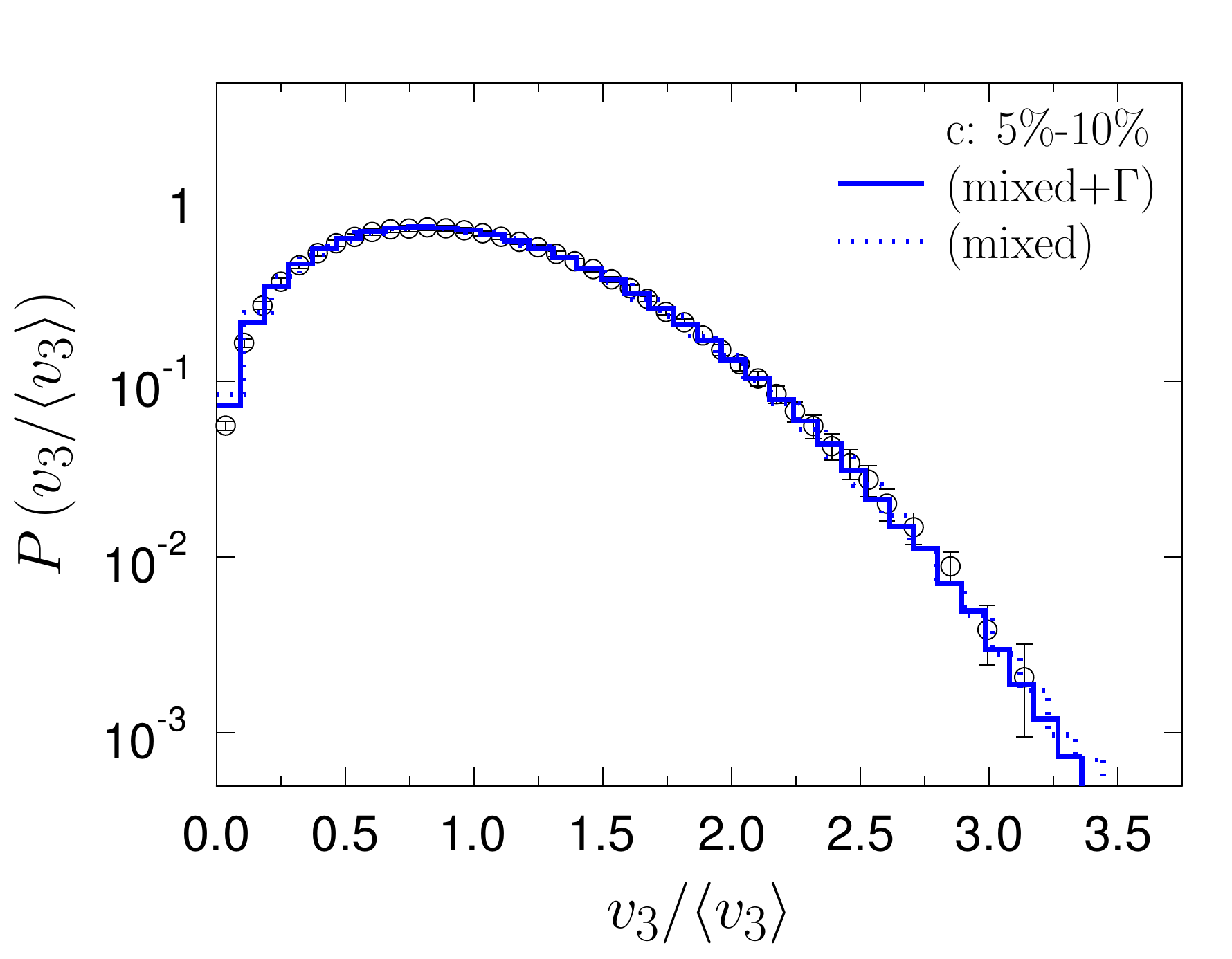}
\includegraphics[width=0.51\textwidth]{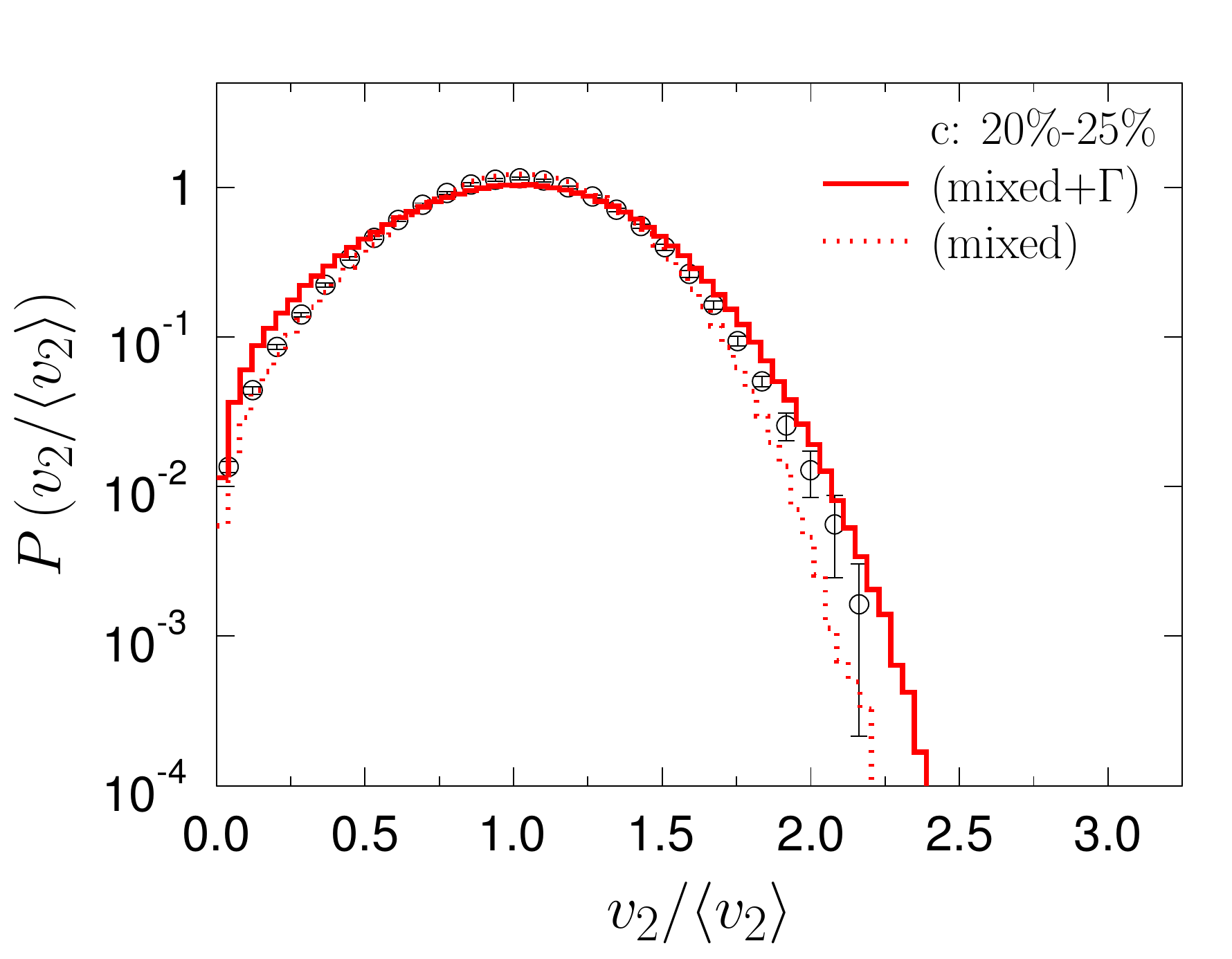}\includegraphics[width=0.51\textwidth]{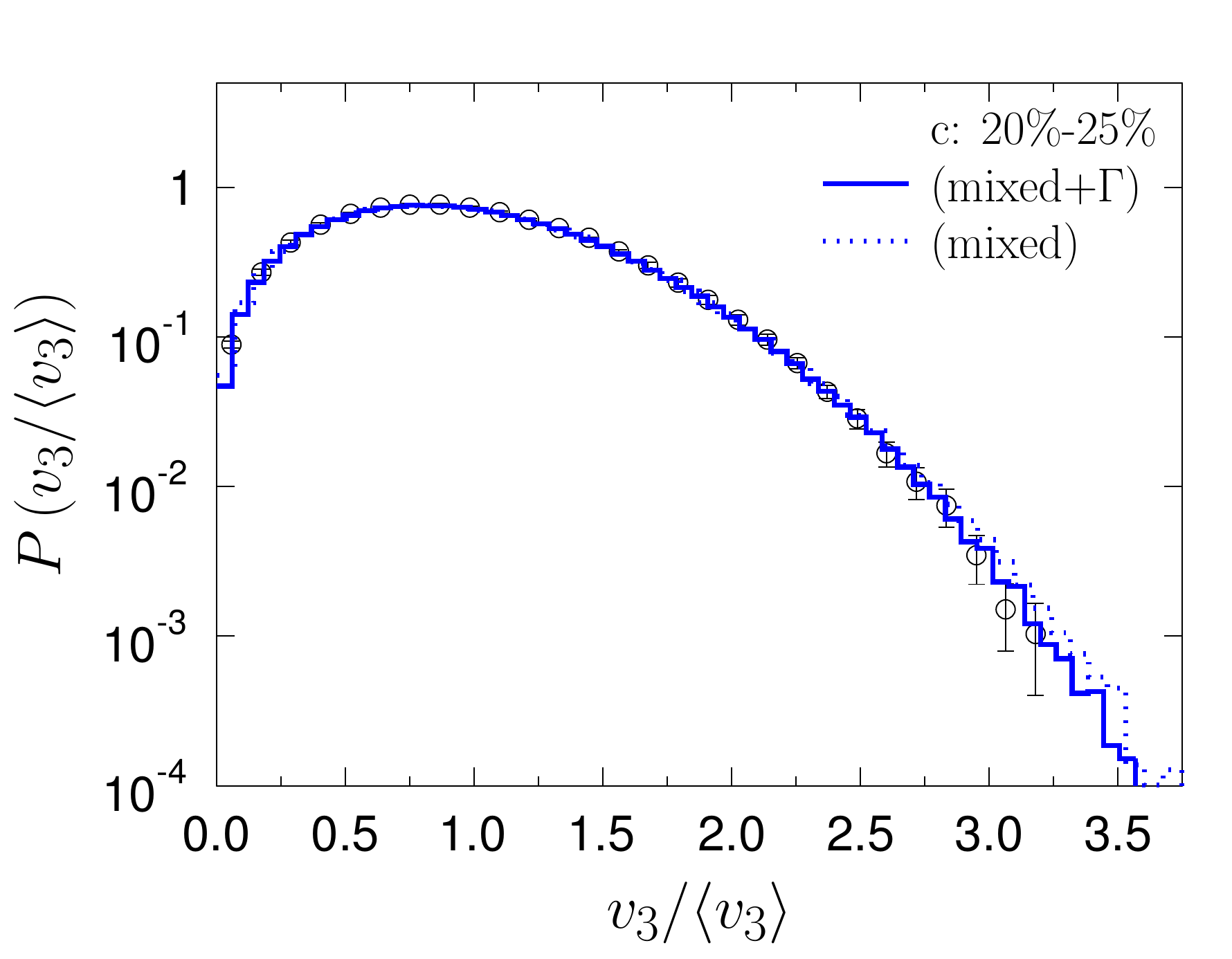}
\includegraphics[width=0.51\textwidth]{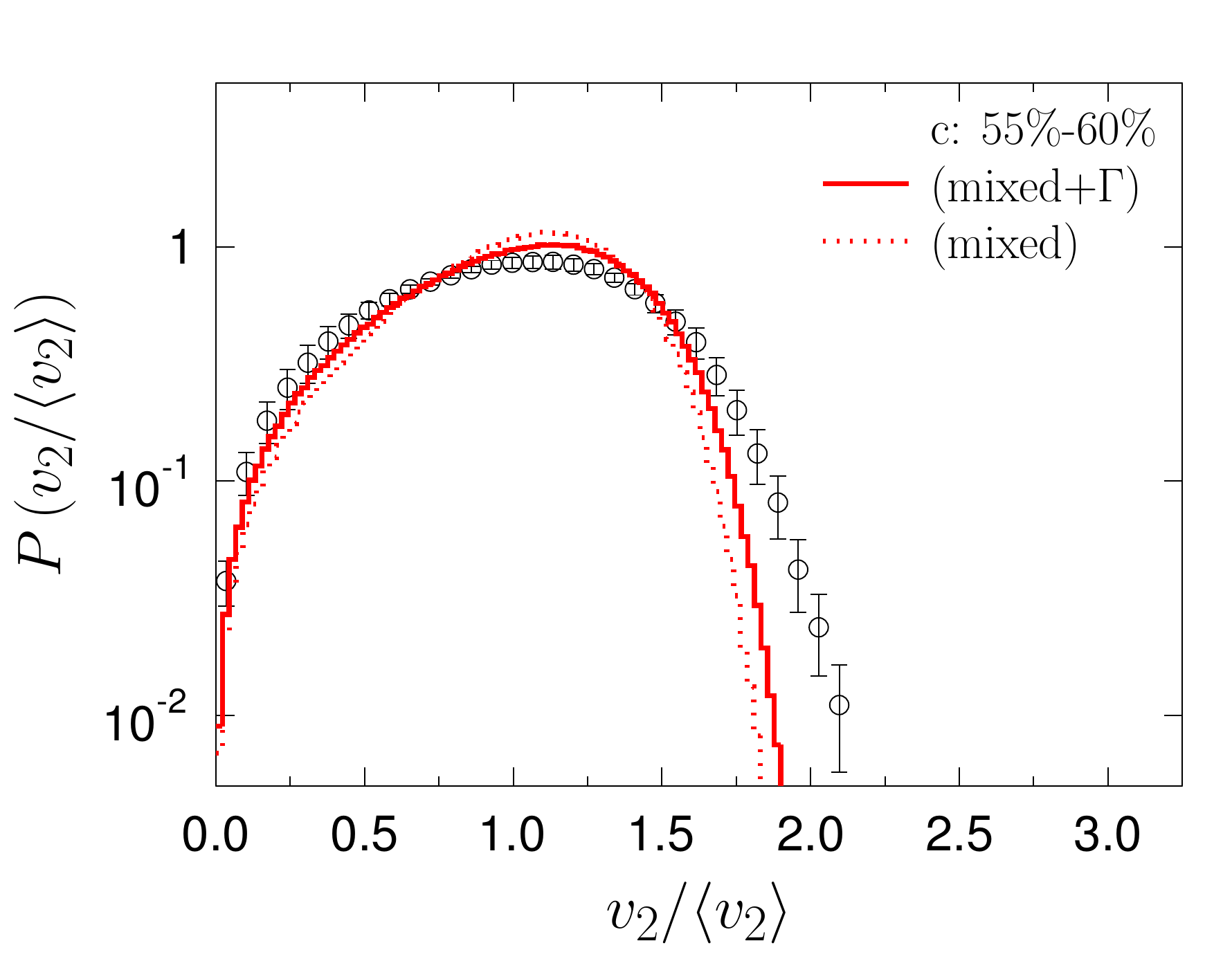}\includegraphics[width=0.51\textwidth]{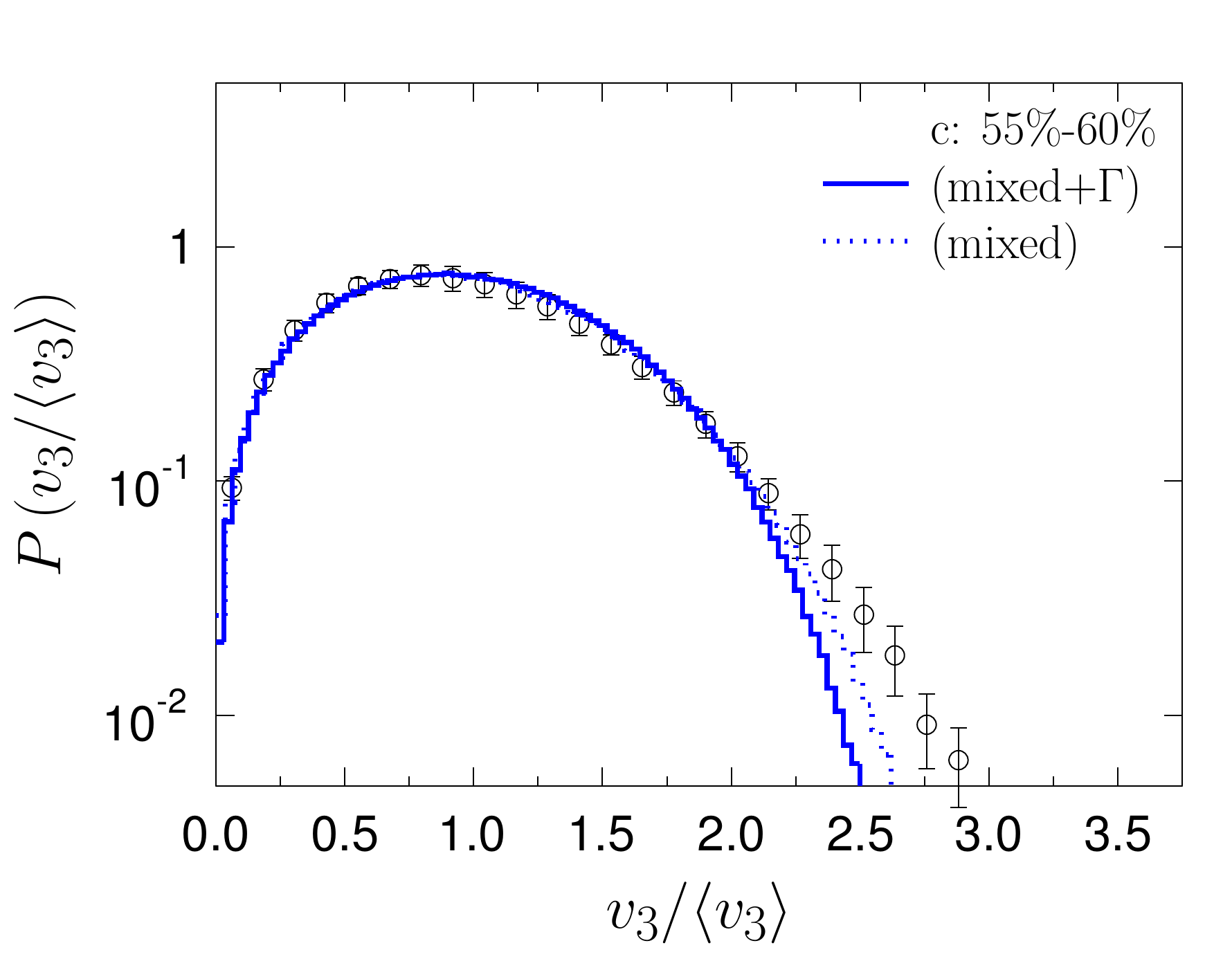}
\end{center}
\vspace{-4mm}
\caption{Distributions of $\epsilon_n/\langle \epsilon_n \rangle$ for the model calculations, compared to the experimental distribution of  
$v_n/\langle v_n \rangle$ from the ATLAS collaboration~\cite{Aad:2013xma}. Top row: centrality $5-10\%$,  middle row: centrality $20-25\%$,  
bottom row: centrality $55-60\%$.
\label{Fig:dist}}
\end{figure}

From Eq.~(\ref{eq:linear}) one obtains immediately the relation for the scaled (i.e., independent of the mean) quantities 
\begin{eqnarray}
\frac{v_n}{\langle v_n \rangle}=\frac{\epsilon_n}{\langle \epsilon_n \rangle}, \;\;n=2,3, \label{eq:linear2}
\end{eqnarray}
where $\langle . \rangle$ denotes averaging over events in the given class. 
Equation (\ref{eq:linear2}) means that the event-by-event 
distributions of the scaled quantities should be equal, i.e., $p({v_n}/{\langle v_n \rangle})=p({\epsilon_n}/{\langle \epsilon_n \rangle})$. 
As seen from Fig.~\ref{Fig:dist}, this is indeed the case to expected accuracy. 
The agreement with the ATLAS data is not perfect, especially for the ellipticity case, where the model distribution is somewhat too wide for the most central events and to narrow for the most peripheral events. This agrees qualitatively with the results of Refs.~\cite{Renk:2014jja,Fu:2015wba}.

\begin{figure}[tb]
\centerline{%
\includegraphics[width=0.7\textwidth]{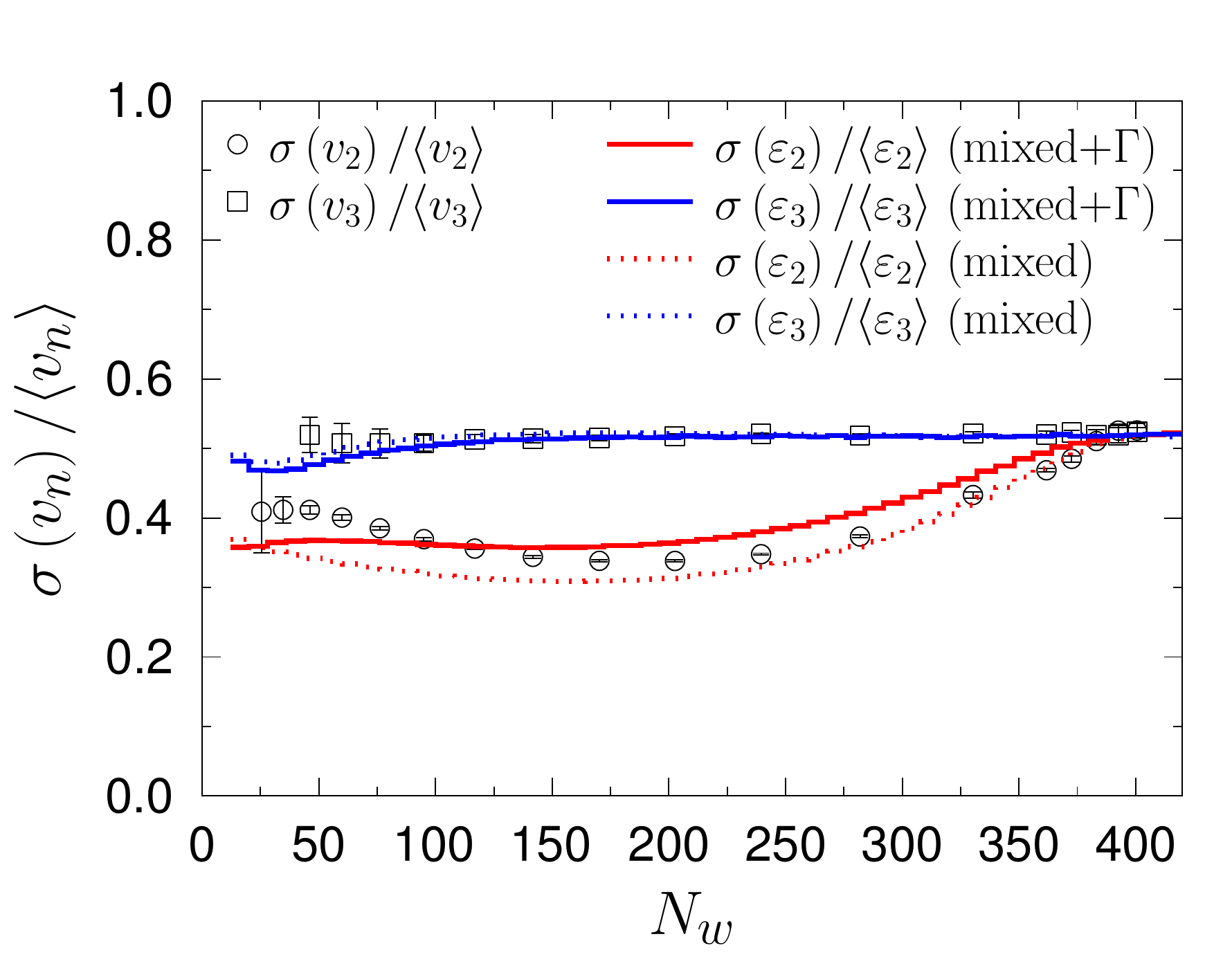}}
\caption{The scaled event-by-event standard deviation for the eccentricities, $\sigma(\epsilon_n)/\langle \epsilon_n \rangle$, and for the harmonic flow coefficients $\sigma(v_n)/\langle v_n \rangle$, plotted as functions of the number of wounded nucleons $N_{w}$. The dashed (solid) lines correspond to our simulation in the mixed (mixed+$\Gamma$). The data come from the ATLAS~\cite{Aad:2013xma} collaboration. \label{Fig:sc_sigma}}
\end{figure}

\begin{figure}[tb]
\centerline{\includegraphics[width=0.7\textwidth]{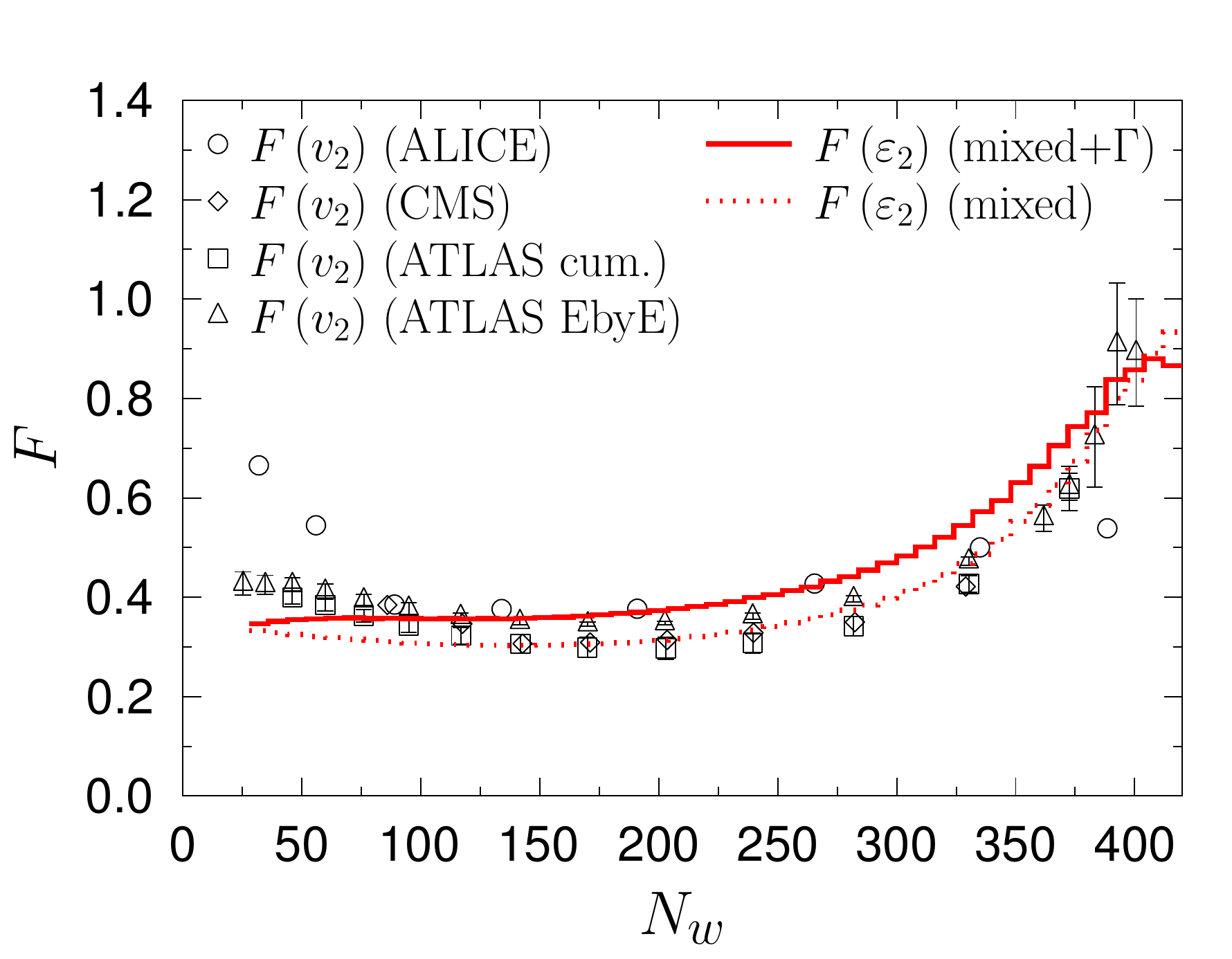}}
\caption{The relative elliptic flow event-by-event fluctuations measure $F\left(v_{2}\right)$, plotted as a function of the number of wounded nucleons $N_{w}$. Result of our simulation with the mixed+$\Gamma$ model is displayed with the solid line, whereas the dashed line shows the outcome of the mixed model. 
The points show the data from ATLAS~\cite{Aad:2014vba}, ALICE~\cite{Aamodt:2010pa}, and CMS~\cite{Chatrchyan:2013kba} collaborations.
\label{Fig:f2}}
\end{figure}

\begin{figure}[tb]
\centerline{\includegraphics[width=0.7\textwidth]{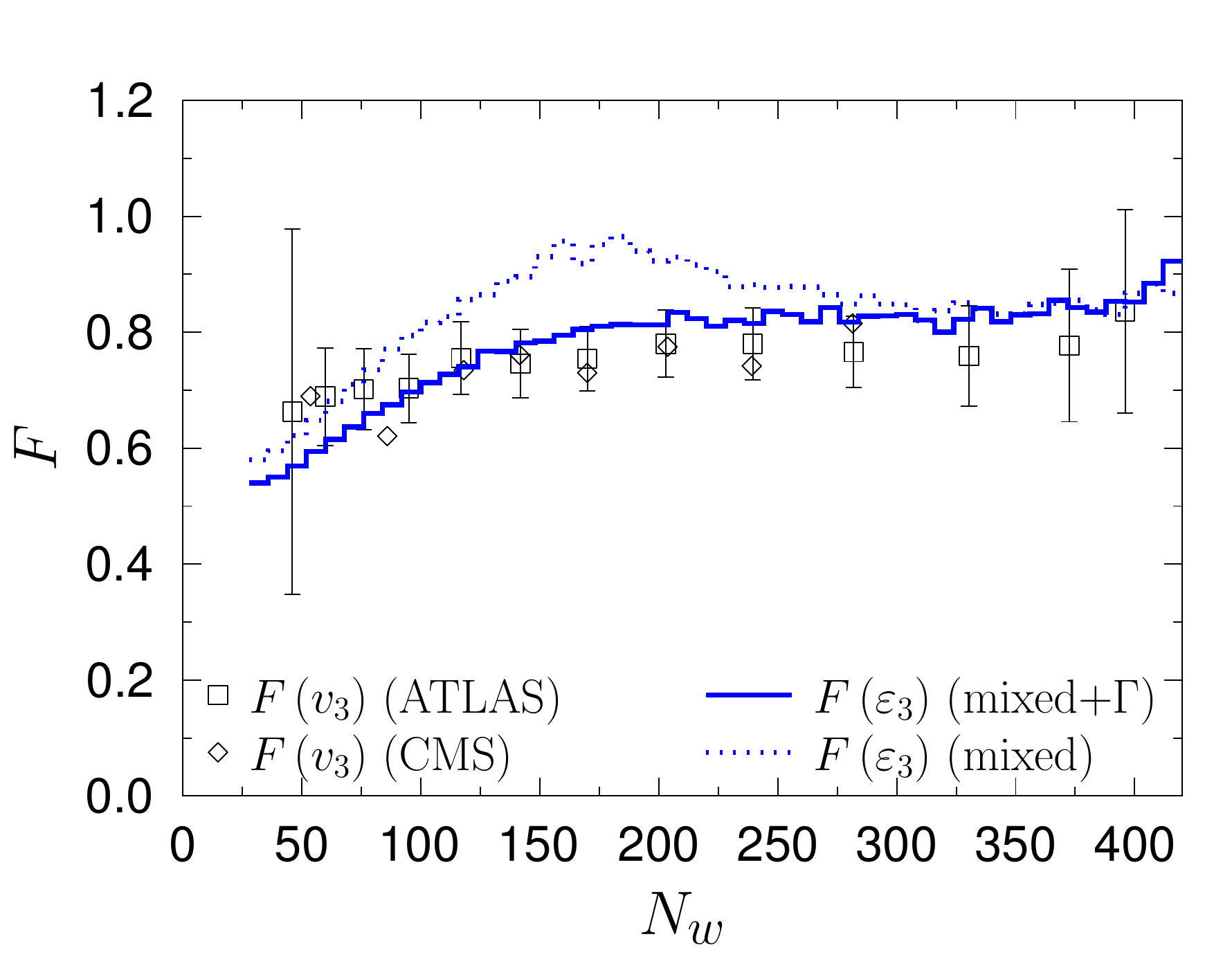}}
\caption{Same as in Fig.~\ref{Fig:f2} but for $F\left(v_{3}\right)$. \label{Fig:f3}}
\end{figure}

With the approximate equality of the distributions for the scaled quantities, the same feature holds for various statistical moments. 
Below, we explore the two-particle and four-particle cumulants moments~\cite{Borghini:2000sa}, defined as
\begin{eqnarray}
     \epsilon_n\{2\} &=&  \langle \epsilon_2^2 \rangle^{1/2}, 
\nonumber \\
     \epsilon_n\{4\} &=& 2 \left ( \langle \epsilon_n^2 \rangle^2 -
\langle \epsilon_n^4 \rangle \right )^{1/4}. \label{eq:cumul}
\end{eqnarray}
More specifically, we take the scaled event-by event standard deviation, 
${\sigma(\epsilon_n)}/{\langle \epsilon_n \rangle}$, and the $F_n(\epsilon_n)$ moments
defined as 
\begin{equation}
F\left(\epsilon_{n}\right)=\sqrt{\frac{\epsilon_{n}\{2\}^{2}-\epsilon_{n}\{4\}^{2}}{\epsilon_{n}\{2\}^{2}+\epsilon_{n}\{4\}^{2}}}.
\end{equation}
These measures are analogously defined for the flow coefficients $v_n$.
According to what has been said, one expects the approximate relations
\begin{eqnarray}
 \frac{\sigma(\epsilon_n)}{\langle \epsilon_n \rangle} \simeq \frac{\sigma(v_n)}{\langle v_n \rangle}, \;\; n=2,3
\end{eqnarray}
and 
\begin{eqnarray}
F(\epsilon_n) \simeq F(v_n), \;\; n=2,3.
\end{eqnarray}

The comparison with the recent LHC data is made in Figs.~\ref{Fig:sc_sigma}-\ref{Fig:f3}, where we plot the  scaled standard deviation and the $F_n$ as functions of the number of 
wounded nucleons $N_w$. We note a very reasonable agreement for sufficiently central collisions ($N_w>100$). These results should be juxtaposed to Figs.~13-14 of Ref.~\cite{Aad:2014vba} or Fig.~18 from Ref.~\cite{Aad:2013xma}, which show that these papers report incorrect results from the Glauber simulations. We note that for the most central events, where only fluctuations contribute to eccentricities, we have $\sigma(\epsilon_n)/\langle \epsilon_n \rangle \to \sqrt{4/\pi-1}$~\cite{Broniowski:2007ft}, and $F(\epsilon_n)\to 1$. For peripheral collisions ($N_w<100$) the agreement is poorer, calling for 
improvement. 

\section{Quadrangular flow}

\begin{figure}[tb]
\centerline{\includegraphics[width=0.7\textwidth]{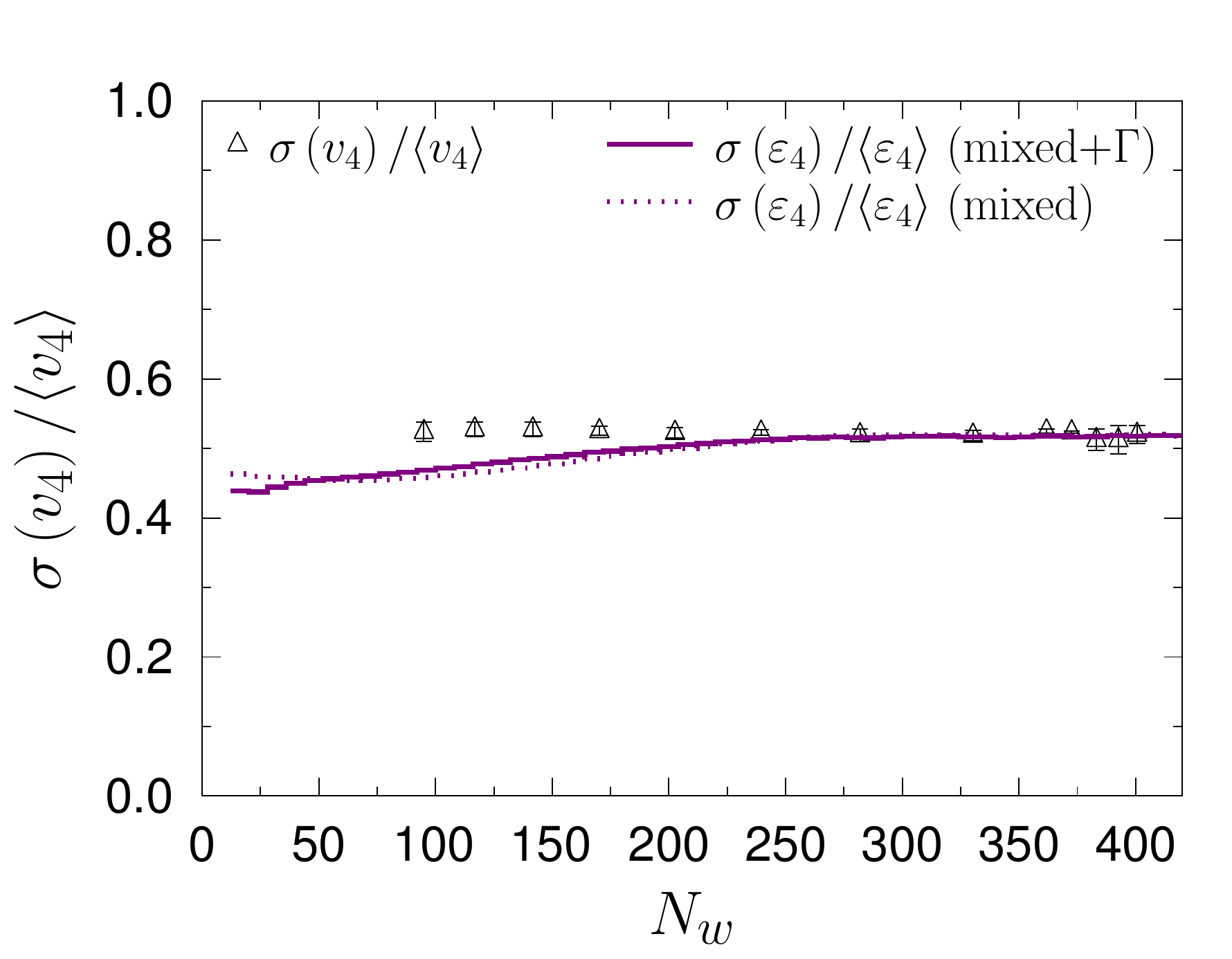}} \vspace{-7mm}
\centerline{\includegraphics[width=0.7\textwidth]{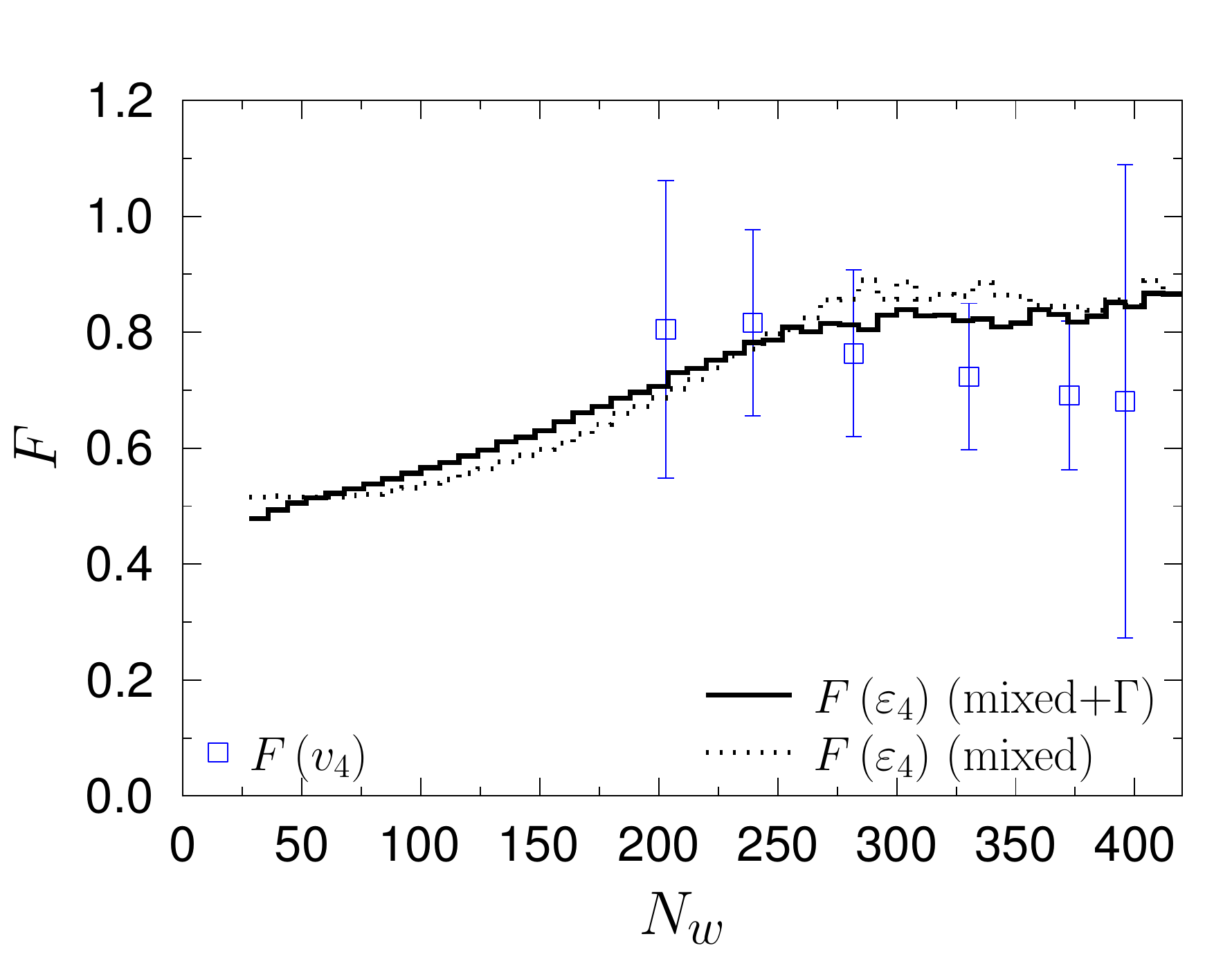}}
\caption{The measures $\sigma(\epsilon_4)/\langle \epsilon_4 \rangle$ (top panel) and $F\left(v_{4}\right)$ (bottom panel), compared to the ATLAS data~\cite{Aad:2013xma} 
for the corresponding quantities for the quadrangular flow coefficient $v_4$. \label{Fig:f4}}
\end{figure}

The above analysis was carried out for $n=2$ and $n=3$, as it has been claimed in the literature that higher rank flow coefficients are more complicated due to non-linear effects, 
incorporating for instance the $\epsilon_2^2$ contributions in $v_4$, etc.~\cite{Teaney:2010vd}. Nevertheless, we have tested that taking the relation (\ref{eq:linear}) also for the case $n=4$, i.e. 
$v_4=\kappa \epsilon_4$,
leads to very reasonable behavior of the flow fluctuations. The results are shown in Fig.~\ref{Fig:f4}, which are fine for central and semi-central events. On the other hand, taking the 
strong nonlinear 
response $v_4=\kappa \epsilon_2^2$ would lead to substantial disagreement, with $\sigma(\epsilon_2)/\langle \epsilon_2 \rangle \to 1$, high above the data for $\sigma(v_4)/\langle v_4 \rangle$. 
We note that a small nonlinear admixture in $v_4$ is not excluded, but the bulk contribution should come from just the linear response to $\epsilon_4$, as 
suggested by Fig.~\ref{Fig:f4}.

\section{Conclusions}

Our main results are as follows:

\begin{enumerate}
 \item Glauber model works within expected accuracy for the flow measures $\sigma(\epsilon_n)/\langle \epsilon_n \rangle$ and $F\left(v_{n}\right)$, 
          for $n=2,3$, but also for $n=4$.
 \item Our results for ellipticity and triangularity fluctuations agree qualitatively with the studies of Ref.~\cite{Niemi:2012aj,Renk:2014jja,Fu:2015wba}, but correct the results of the Glauber simulations presented in Refs.~\cite{Aad:2013xma,Aad:2014vba}.
 \item Our results for the investigated measures do not depend strongly on the details of the Glauber model (overlaid distribution, correlations in the nuclear distributions,
 wounding profile, etc.), hence are robust for the investigation of flow fluctuations.
\end{enumerate}

\bigskip
\bigskip 

We thank Piotr Bo\.zek for a helpful discussion. Research supported by the National Science Center grants DEC-2012/05/B/ST2/02528 and DEC-2012/06/A/ST2/00390. 

\appendix

\section{Running the simulations}

To reproduce the results of the simulations presented in this paper, or to extend them to other physical cases, the user should
download the package {\tt GLISSANDO~2}~ver.~2.9~\cite{Rybczynski:2013yba} from the web page 

\begin{verbatim}
http://www.ujk.edu.pl/homepages/mryb/GLISSANDO/
\end{verbatim}
\noindent and after unpacking execute (on UNIX systems) the following commands:

{\small
\begin{verbatim}
make
./glissando2 input/mixed_gamma.dat output/mixed_gamma_01.root
root -b -l -q -x "macro/eps_fluct.C(\"output/mixed_gamma\",1)"
\end{verbatim}
}
\noindent More statistics can be accumulated by running, for instance
{\small
\begin{verbatim}
./glissando2 input/mixed_gamma.dat output/mixed_gamma_02.root
...
./glissando2 input/mixed_gamma.dat output/mixed_gamma_10.root
root -b -l -q -x "macro/eps_fluct.C(\"output/mixed_gamma\",10)"
\end{verbatim}
}
\noindent The plots are placed in the {\tt output} directory. The present code has been checked with {\tt ROOT} ver.~5.34.

\bibliography{hydr}

\begin{thebibliography}{10}

\bibitem{Sorensen:2006nw}
STAR, P. Sorensen,
\newblock J. Phys. G34 (2007) S897, nucl-ex/0612021.

\bibitem{Alver:2007zz}
PHOBOS, B. Alver,
\newblock Int. J. Mod. Phys. E16 (2007) 3331.

\bibitem{Alver:2007rm}
PHOBOS, B. Alver et~al.,
\newblock J. Phys. G34 (2007) S907, nucl-ex/0701049.

\bibitem{Aamodt:2010pa}
ALICE Collaboration, K. Aamodt et~al.,
\newblock Phys. Rev. Lett. 105 (2010) 252302, 1011.3914.

\bibitem{Abelev:2012di}
ALICE, B. Abelev et~al.,
\newblock Phys. Lett. B719 (2013) 18, 1205.5761.

\bibitem{Aad:2013xma}
ATLAS Collaboration, G. Aad et~al.,
\newblock JHEP 1311 (2013) 183, 1305.2942.

\bibitem{Aad:2014vba}
ATLAS, G. Aad et~al.,
\newblock Eur. Phys. J. C74 (2014) 3157, 1408.4342.

\bibitem{Chatrchyan:2013kba}
CMS, S. Chatrchyan et~al.,
\newblock Phys. Rev. C89 (2014) 044906, 1310.8651.

\bibitem{Luzum:2013yya}
M. Luzum and H. Petersen,
\newblock J. Phys. G41 (2014) 063102, 1312.5503.

\bibitem{Qiu:2011iv}
Z. Qiu and U.W. Heinz,
\newblock Phys. Rev. C84 (2011) 024911, 1104.0650.

\bibitem{Niemi:2012aj}
H. Niemi et~al.,
\newblock Phys. Rev. C87 (2013) 054901, 1212.1008.

\bibitem{Bzdak:2013rya}
A. Bzdak, P. Bo\.zek and L. McLerran,
\newblock Nucl.Phys. A927 (2014) 15, 1311.7325.

\bibitem{Renk:2014jja}
T. Renk and H. Niemi,
\newblock Phys. Rev. C89 (2014) 064907, 1401.2069.

\bibitem{Fu:2015wba}
J. Fu,
\newblock Phys. Rev. C92 (2015) 024904.

\bibitem{Bravina:2015sda}
L.V. Bravina et~al.,
\newblock (2015), 1509.02692.

\bibitem{Broniowski:2007nz}
W. Broniowski, M. Rybczy\'nski and P. Bo\.zek,
\newblock Comput. Phys. Commun. 180 (2009) 69, 0710.5731.

\bibitem{Rybczynski:2013yba}
M. Rybczy\'nski et~al.,
\newblock Comput. Phys. Commun. 185 (2014) 1759, 1310.5475.

\bibitem{Broniowski:2007ft}
W. Broniowski, P. Bo\.zek and M. Rybczy\'nski,
\newblock Phys. Rev. C76 (2007) 054905, 0706.4266.

\bibitem{Bialas:1976ed}
A. Bia\l{}as, M. B\l{}eszy\'nski and W. Czy\.z,
\newblock Nucl. Phys. B111 (1976) 461.

\bibitem{Kharzeev:2000ph}
D. Kharzeev and M. Nardi,
\newblock Phys. Lett. B507 (2001) 121, nucl-th/0012025.

\bibitem{SchaffnerBielich:2001qj}
J. Schaffner-Bielich et~al.,
\newblock Nucl. Phys. A705 (2002) 494, nucl-th/0108048.

\bibitem{Back:2001xy}
PHOBOS, B.B. Back et~al.,
\newblock Phys. Rev. C65 (2002) 031901, nucl-ex/0105011.

\bibitem{Back:2004dy}
PHOBOS, B.B. Back et~al.,
\newblock Phys. Rev. C70 (2004) 021902, nucl-ex/0405027.

\bibitem{Bozek:2013uha}
P. Bo\.zek and W. Broniowski,
\newblock Phys. Rev. C88 (2013) 014903, 1304.3044.

\bibitem{Olszewski:2013qwa}
A. Olszewski and W. Broniowski,
\newblock Phys.Rev. C88 (2013) 044913, 1303.5280.

\bibitem{cmswiki}
CMS, S. Chatrchyan et~al.,
\newblock CMSPublic Web  (2012),
\newblock
  {http://twiki.cern.ch/twiki/bin/view/CMSPublic/PhysicsResultsHIN12015}.

\bibitem{Rybczynski:2013mla}
M. Rybczy\'nski and Z. W\l{}odarczyk,
\newblock J.Phys. G41 (2013) 015106, 1307.0636.

\bibitem{Alvioli:2009ab}
M. Alvioli, H.J. Drescher and M. Strikman,
\newblock Phys. Lett. B680 (2009) 225, 0905.2670.

\bibitem{Ollitrault:1992bk}
J.Y. Ollitrault,
\newblock Phys. Rev. D46 (1992) 229.

\bibitem{Alver:2006wh}
PHOBOS Collaboration, B. Alver et~al.,
\newblock Phys. Rev. Lett. 98 (2007) 242302, nucl-ex/0610037.

\bibitem{Voloshin:2006gz}
S.A. Voloshin,
\newblock (2006), nucl-th/0606022.

\bibitem{Hama:2007dq}
Y. Hama et~al.,
\newblock Phys.Atom.Nucl. 71 (2008) 1558, 0711.4544.

\bibitem{Alver:2010gr}
B. Alver and G. Roland,
\newblock Phys. Rev. C81 (2010) 054905, 1003.0194.

\bibitem{Luzum:2011mm}
M. Luzum,
\newblock J.Phys. G38 (2011) 124026, 1107.0592.

\bibitem{Bhalerao:2014xra}
R.S. Bhalerao, J.Y. Ollitrault and S. Pal,
\newblock Phys. Lett. B742 (2015) 94, 1411.5160.

\bibitem{Bozek:2015bna}
P. Bo\.zek and W. Broniowski,
\newblock (2015), 1506.02817.

\bibitem{Bozek:2010vz}
P. Bo\.zek, W. Broniowski and J. Moreira,
\newblock Phys. Rev. C83 (2011) 034911, 1011.3354.

\bibitem{Bozek:2014cva}
P. Bo{\.z}ek et~al.,
\newblock Phys.Rev. C90 (2014) 064902, 1410.7434.

\bibitem{Teaney:2010vd}
D. Teaney and L. Yan,
\newblock Phys. Rev. C83 (2011) 064904, 1010.1876.

\bibitem{Borghini:2000sa}
N. Borghini, P.M. Dinh and J.Y. Ollitrault,
\newblock Phys. Rev. C63 (2001) 054906, nucl-th/0007063.

\end{thebibliography}

\end{document}